\documentclass[sigconf,nonacm=true]{acmart}

\copyrightyear{2020}
\acmYear{2020}
\setcopyright{acmlicensed}\acmConference[ACSAC 2020]{Annual Computer Security Applications Conference}{December 7--11, 2020}{Austin, USA}
\acmBooktitle{Annual Computer Security Applications Conference (ACSAC 2020), December 7--11, 2020, Austin, USA}
\acmPrice{15.00}
\acmDOI{10.1145/3427228.3427291}
\acmISBN{978-1-4503-8858-0/20/12}

\author{Dmitry Belyavsky}
\affiliation{
\institution{Cryptocom Ltd.}
\city{Moscow}
\country{Russian Federation}
}
\email{beldmit@cryptocom.ru}

\author{Billy Bob Brumley}
\orcid{0000-0001-9160-0463}
\affiliation{
\institution{Tampere University}
\city{Tampere}
\country{Finland}
}
\email{billy.brumley@tuni.fi}

\author{Jes\'us-Javier Chi-Dom\'inguez}
\orcid{0000-0002-9753-7263}
\affiliation{
\institution{Tampere University}
\city{Tampere}
\country{Finland}
}
\email{jesus.chidominguez@tuni.fi}

\author{Luis Rivera-Zamarripa}
\orcid{0000-0002-1779-421X}
\affiliation{
\institution{Tampere University}
\city{Tampere}
\country{Finland}
}
\email{luis.riverazamarripa@tuni.fi}

\author{Igor Ustinov}
\affiliation{
\institution{Cryptocom Ltd.}
\city{Moscow}
\country{Russian Federation}
}
\email{igus@cryptocom.ru}

\usepackage[T1]{fontenc}
\usepackage{graphicx}
\usepackage{xspace}
\usepackage{lipsum}
\usepackage{multirow}

\newcommand{\footurl}[1]{\footnote{\url{#1}}}
\newcommand{\CVE}[1]{\href{https://cve.mitre.org/cgi-bin/cvename.cgi?name=CVE-#1}{\mbox{CVE-#1}}}
\newcommand{\rfc}[1]{\href{https://tools.ietf.org/html/rfc#1}{RFC #1}~\citep{rfc:#1}}
\newcommand{\opensslpr}[1]{\href{https://github.com/openssl/openssl/pull/#1}{\mbox{PR \##1}}}
\newcommand{\code}[1]{\texttt{#1}\xspace}
\newcommand{\curve}[1]{\code{#1}}
\newcommand{\gostengine}{\code{gost\discretionary{-}{}{-}engine}}
\newcommand{\Paragraph}[1]{\subsubsection*{#1}}

\newcommand{\KEYWORDS}{%
applied cryptography;
public key cryptography;
elliptic curve cryptography;
software engineering;
software testing;
formal verification;
GOST;
NSS;
OpenSSL}

\title{Set It and Forget It! Turnkey ECC for Instant Integration}

\begin{abstract}
Historically, Elliptic Curve Cryptography (ECC) is an active field of applied
cryptography where recent focus is on high speed, constant time, and formally
verified implementations. While there are a handful of outliers where all these
concepts join and land in real-world deployments, these are generally on a
case-by-case basis: e.g.\ a library may feature such X25519 or P-256 code, but
not for all curves. In this work, we propose and implement a methodology that
fully automates the implementation, testing, and integration of ECC stacks with
the above properties. We demonstrate the flexibility and applicability of our
methodology by seamlessly integrating into three real-world projects: OpenSSL,
Mozilla's NSS, and the GOST OpenSSL Engine, achieving roughly 9.5x, 4.5x, 13.3x, and 3.7x 
speedup on any given curve for key generation, key agreement, signing, and verifying, respectively.
Furthermore, we showcase the efficacy of our testing methodology by uncovering
flaws and vulnerabilities in OpenSSL, and a specification-level vulnerability in
a Russian standard. Our work bridges the gap between significant applied
cryptography research results and deployed software, fully automating the
process.
 \end{abstract}

\begin{document}

\maketitle

\section{Introduction} \label{sec:intro}

In 1976, Whitfield Diffie and Martin Hellman published the first key-exchange protocol~\cite{DBLP:journals/tit/DiffieH76}
(based on Galois field arithmetic) that provides the capability for two different users to agree upon
a shared secret between them.
In 1985, \citet{DBLP:conf/crypto/Miller85} and \citet{DBLP:conf/crypto/Koblitz90} proposed public-key
cryptosystems based on the group structure of an elliptic curve over Galois fields; from these works,
an Elliptic Curve Diffie-Hellman (ECDH) variant arose. In 1994, Scott Vanstone proposed an Elliptic Curve Digital
Signature Algorithm (ECDSA) variant (for more details see~\cite{DBLP:journals/ijisec/JohnsonMV01}).
However, the main advantage of using Elliptic Curve Cryptography
(ECC) is the smaller keys compared to their Galois field DH and DSA initial proposals.

From the birth of ECC, which was focused on its mathematical description, the
study, analysis, and improvement of elliptic curve arithmetic to achieve
performant, constant time, {\it exception-free}, and formally verified ECC
implementations are clear research trends. Nevertheless, practice sometimes
misaligns with theory, and by integrating theoretic works into real-world
deployments, vulnerabilities arise and compromise the given ECC scheme security.

\Paragraph{Motivation}
On the practice side, there is no shortage of examples of this misalignment.
\citet{DBLP:conf/asiacrypt/BrumleyH09} published the first (microarchitecture)
timing attack on OpenSSL's ECC implementation in 2009, with countermeasures by
\citet{DBLP:conf/fc/Kasper11} and later \citet{DBLP:journals/jce/GueronK15}. But
OpenSSL supports over 80 named curves, and the scope of these countermeasures is
only three: NIST curves P-224, P-256, and P-521, even later augmented with
formal verification guarantees \cite{DBLP:conf/concur/PolyakovTWY18} after
patching defects \cite{DBLP:conf/kbse/LiuSTWY19}. \CVE{2018-5407} ``PortSmash''
\cite{DBLP:conf/sp/AldayaBHGT19} finally led to wider countermeasures
\cite{DBLP:conf/acsac/TuveriHGB18} a decade later, but small leakage  persists
in the recent ``LadderLeak'' attack \cite{temp:ladderleak}. Still, even if
current solutions hedge against timing attacks, the question of functional
correctness remains: \CVE{2011-4354} from \cite{DBLP:conf/ctrsa/BrumleyBPV12} is
the only real-world bug attack \cite{DBLP:journals/joc/BihamCS16} we are aware
of, deterministically recovering P-256 keys remotely by exploiting a carry
propagation defect.

BoringSSL approaches the constant time and functional correctness issues by
narrowing features, only supporting P-224, P-256, and X25519, leveraging formal
verification guarantees for Galois field arithmetic from
Fiat \cite{DBLP:conf/sp/ErbsenPGSC19}. Mozilla's NSS approach is similar,
removing support for the vast majority of curves---two of which (P-256, X25519)
leverage the formal verification results from
HACL$^*$ \cite{DBLP:conf/ccs/ZinzindohoueBPB17}, while others still use generic
legacy code with no protections or guarantees. Stripping support is not a viable
option for fuller-featured libraries, OpenSSL being one example but generally
any project with even a slightly larger scope. \textit{How can these projects
retain features yet provide constant time and functional correctness
confidence?}

\Paragraph{Contributions}
Our main contribution focuses on fully automatic implementation, testing, and
integration of ECC stacks on real-world projects like OpenSSL, Mozilla's NSS,
and GOST OpenSSL Engine. Our full-stack ECC implementations achieve about 
9.5x, 4.5x, 13.3x, and 3.7x speedup for key generation, key agreement, signing, and verifying, respectively. 
Furthermore, our flexible and applicable proposal can be easily adapted to any curve model. 
To our knowledge, this is the first hybrid ECC implementation between {\it short
Weierstrass} and {\it Twisted Edwards} curves, which has been integrated to
OpenSSL. Additionally, our methodology allowed us to find and fix very special
vulnerabilities on development versions of OpenSSL and official Russian
standards for cryptography.

\Paragraph{Outline}
\autoref{sec:background} gives the elliptic curve background concepts related to
curve models and cryptosystems; in particular,
\autoref{sec:gostmath} describes GOST and the related OpenSSL GOST Engine.
\autoref{sec:ecckat} introduces our library-agnostic unit and regression testing
framework for ECC implementations (ECCKAT); while \autoref{sec:ecckiila}
presents our dynamic ECC layer generation (ECCKiila) and performance results.
Finally, \autoref{sec:conclusion} draws conclusions.
\section{Background} \label{sec:background}
An elliptic curve \(E\) defined over a Galois field \(GF(p)\), is usually described
by an equation of the following form
\begin{align}
E \colon y^2 = x^3 + ax + b,\;\;\;a\in GF(p),\;b \in GF(p);
\label{sec:background:eq_ec}
\end{align}
called a short Weierstrass curve. Furthermore, a point on the
curve \(E\) is a pair \((x,y)\) satisfying \eqref{sec:background:eq_ec}, but there is
also a {\it point at infinity} denoted \(\mathcal{O}\), which plays the role of the
neutral element on \(E\). Additionally, given a positive integer \(k\), point multiplication
is the computation of \(k\) times a given point \(P\) denoted by \([k]P\).
The order of a point \(P\) on \(E\) corresponds with the smallest positive integer \(q\) such
that \([q]P\) gives \(\mathcal{O}\). In our work, we assume the cardinality
of \(E\) is equal to \(h\cdot q\) where \(q\) is a prime number with $\lg(q) \approx \lg(p)$,
and \(h \in \{1,4,8\}\).

When $4$ divides $h$, there is a {\it Twisted Edwards curve}
\begin{align}
E_t \colon eu^2 + v^2 = 1 + du^2v^2
\label{sec:background:eq_ed}
\end{align}
having the same cardinality,
and each point on \(E_w\) \eqref{sec:background:eq_ec} maps into \(E_t\), and vice versa, using the mappings
\begin{align}
\label{sec:ecckiila:eq2}
(x,y)&\mapsto (u,v) \coloneqq \left(\frac{x - t}{y}, \frac{x - t - s}{x - t + s}\right), \text{ and}\\
\label{sec:ecckiila:eq1}
(u,v)&\mapsto (x,y) \coloneqq \left(\frac{s(1 + v)}{1 - v} + t, s\frac{(1 + v)}{(1 - v)u}\right),
\end{align}
where \(s = (e-d)/4 \bmod p\), \(t = (e+d)/6 \bmod p\), \(a = (s^2 - 3t^2) \bmod p\), and \(b = (2t^3 - t\cdot s^2) \bmod p\).

\Paragraph{Projective points on the short Weierstrass curve}
We choose to work with {\it projective} points \((X:Y:Z)\)
satisfying \(ZY^2 = X^3 + aXZ^2 + bZ^3\) where the {\it affine} point \(\left(X/Z,Y/Z\right)\)
belongs to \(E_w\). Moreover, the projective representation of \(\mathcal{O}\) is \((0:1:0)\),
which does not satisfy the {\it affine} curve equation of \(E_w\).

Because of the nature of the short Weierstrass curves, one needs to handle some
exceptions when: (i) adding or doubling points with \(\mathcal{O}\); (ii) adding
points $P + Q$ when $P = \pm Q$. In particular, any {\it mixed} point addition takes as
inputs a {\it projective} point and an {\it affine} point, which implies no
exception-free implementation will be possible for this {\it mixed} point
addition---\(\mathcal{O}\) has no {\it affine} representation!

Failure to use {\it exception-free} formulas could lead to successful
\textit{exceptional procedure attacks}~\cite{DBLP:conf/pkc/IzuT03}, implying a
possible break of ECC security. Still, apart from theoretical attacks there is
the question of functional correctness. For example, \CVE{2017-7781} affected
Mozilla's NSS, failing to account for the $P = \pm Q$ exceptions in textbook
mixed Jacobian-affine point addition---a bug present in their codebase for over
a decade.

\Paragraph{Projective points on the Twisted Edwards curve}
To achieve efficient curve arithmetic, we choose to work with {\it extended projective}
points \((X:Y:T:Z)\) satisfying \(eX^2Z^2 + Y^2Z^2 = Z^4 + dX^2Y^2\), where the {\it affine}
point \(\left(X/Z,Y/Z\right)\) belongs to \(E_t\) and \(T = XY/Z\). The main advantage of
using {\it Twisted Edwards curves} is the ``cheap'' exception-free formula for point
addition; in particular, \((0:1:0:1)\) represents \(\mathcal{O}\) and
corresponds with the {\it affine} point \((0,1)\) on \(E_t\).

The main blocks of ECC cryptosystem implementations consist of (i) key
generation, (ii) key agreement procedure, and (iii) digital signature algorithm.

\Paragraph{Key generation}
Given an order-$q$ point $g$ the user randomly and uniformly chooses a secret key \(\alpha\) from $\{1, \ldots, q-1\}$, and computes the public key \(P = [\alpha]g\).

\Paragraph{Key agreement with cofactor clearing (ECC CDH)}
Assume the users {\it Alice} and {\it Bob} need to agree a secret shared key; thus,
Alice generates her private key \(\alpha_a \in \{1, \ldots, q - 1\}\) and a public key \(P_a = [\alpha_a]g\) by using the key generation block;
similarly, Bob generates \(\alpha_b\) and \(P_b = [\alpha_b]g\).
Next, Alice and Bob compute \(s_{ab} = [h \cdot \alpha_a]P_b\) and \(s_{ba} = [h \cdot \alpha_b]P_a\), respectively.
Consequently,
\[
s_{ab} = [h \cdot \alpha_a]P_b = [h \cdot \alpha_a \cdot \alpha_b]g = [h \cdot \alpha_b \cdot \alpha_a] g = [h \cdot \alpha_b]P_a = s_{ba}
\]
is the secret shared key. The multiplication by $h$ is called \textit{cofactor clearing}
and ensures the protocol fails if $P_a$ or $P_b$ are adversarially in the order-$h$ subgroup.
When $h=1$, ECC CDH~\cite{temp:sp:800-56ARev3} and classical ECDH variants are equivalent.

\Paragraph{Digital signature algorithm (ECDSA)}
The user generates a private key \(\alpha \in \{1, \ldots, q - 1\}\) and a public key \(P = [\alpha]g\) by using the key generation block;
using an approved hash function $\textrm{Hash}()$, the signature $(r,s)$ on message $m$ is computed by
\begin{equation} \label{eq:ecdsa:sign}
r = ([k]g)_x \bmod q, \quad s = k^{-1} (\hat{m} + \alpha r) \bmod q
\end{equation}
where $k$ is a nonce chosen uniformly from $\{1, \ldots, q-1\}$, and $\hat{m}$ denotes the representation of $ \textrm{Hash}(m)$ in $GF(q)$.
The ECDSA signature successfully verifies if \(u_1 = \hat{m} \cdot s^{-1} \bmod q\) and \(u_2 = r \cdot s^{-1} \mod q\) satisfy
\begin{equation} \label{eq:ecdsa:verify}
([u_1]g + [u_2]P)_x = r \bmod q.
\end{equation}
ECDSA is the ECC-equivalent of DSA that instead operates with the multiplicative
group of a Galois field and pre-dates the ECDSA variant by at least a decade.

\Paragraph{Security}
Mathematically speaking, the security of ECC relies on the hardness of computing
an integer \(k\) given \([k]P\) called the \textit{Elliptic Curve Discrete
Logarithm Problem (ECDLP)}.
In certain instances, ECDLP solves
by using the {\it small-subgroup}~\cite{DBLP:conf/crypto/LimL97} when the curve cardinality is smooth, and {\it invalid-curve}~\cite{DBLP:conf/crypto/BiehlMM00}
attacks when the input point \(P\) does not satisfies the curve equation.

As a consequence, ECC implementations often seek to be secure against {\it combined} attacks that use
{\it small-subgroup} attacks with {\it invalid-curve} attacks using the {\it twist} curve \(E'\) determined by the equation
\(y^2 = x^3 + ax - b\). The {\it twist} curve \(E'\) has cardinality
\(h'\cdot q' = p + 1 + t_E\) where \(h\cdot q = p + 1 - t_E\) is the cardinality of
\(E\) and $t_E$ is a curve constant (the \textit{Frobenius trace}).
However, a curve \(E\) is {\it twist secure} if \(h'\) is a small integer and \(q' \approx p\) is a
large prime number. For example, the following two GOST curves are {\it twist secure}:
\begin{itemize}
\item the curve \curve{id\_tc26\_gost\_3410\_2012\_256\_paramSetA} has
\begin{verbatim}
q = 0x3FFFFFFFFFFFFFFFFFFFFFFFFFFFFFFFF027322037
      8499CA3EEA50AA93C9F265,
q'= 0x400000000000000000000000000000000FD8CDDFC8
      7B6635C115AF556C360C67,
\end{verbatim}
and both \(h\) and \(h'\) equal 4;
\item the curve \curve{id\_tc26\_gost\_3410\_2012\_512\_paramSetC} has
\begin{verbatim}
q = 0x400000000000000000000000000000000000000000
      00000000000000000000003673245B9AF954FFB3CC
      5600AEB8AFD33712561858965ED96B9DC310B80FDA
      F7,
q'= 0x3FFFFFFFFFFFFFFFFFFFFFFFFFFFFFFFFFFFFFFFFF
      FFFFFFFFFFFFFFFFFFFFFFC98CDBA46506AB004C33
      A9FF5147502CC8EDA9E7A769A12694623CEF47F023
      ED,
\end{verbatim}
and also both \(h\) and \(h'\) equal 4.
\end{itemize}

\subsection{GOST} \label{sec:gostmath}
The system of Russian cryptographic standards (usually called \textit{GOST
algorithms}) started to develop in the 1980s after decades of \textit{top secret
cryptography}. The first Russian (or, rather, Soviet) relatively open
cryptographic standard was published in 1989, describing symmetric cipher and
MAC algorithms.

The first Russian standard for digital signatures was developed
simultaneously with DSA and these two standards were published in 1994 with
an interval of only four days. Like DSA, the Russian GOST R 34.10-94 was the
ElGamal-style algorithm in Galois field of prime modulo, but the formula was
slightly different: $s=k \hat{m} + \alpha r \bmod q$. The hash function to be used for
calculating $\hat{m}$ was strictly defined and described in a separate standard,
based on the GOST symmetric cipher.

In 2001 the new digital signature standard was adopted---the adaptation of
the previous standard to elliptic curves over $GF(p)$, allowing only
$\lg(p)=256$. The hash function was not changed.

In 2012 the third Russian digital signature standard was adopted, almost
word-to-word copy of the previous standard. The only changes were
(i) the length of $p$ can now be either 256 or 512;
(ii) the standard prescribes to use a new (completely different) hash function.
The official name of the current Russian digital signature standard is GOST R 34.10-2012,
and GOST R 34.11-2012 describes the hash function. The translation of these
standards in English were published as \rfc{7091} and \rfc{6986} respectively.

GOST is utilized in Russia neither everywhere nor only by government agencies,
but in between---including both closed networks and GOST-protected channels over
public networks. For example, businesses use GOST when legally required to protect
transferred data, such as financial sector, medical organizations and so on.

The FIPS-like Russian certification policy divides GOST-im\-ple\-men\-ting software
into two parts.
(i) Various commercial closed-source implementations used mostly in places where
prescribed by Russian regulations. Any certified solution should have a registry
of every instance of their product.
(ii) Open-source implementations (e.g.\ \autoref{sec:gostengine}) are widely used---when
certification is not enforced---for communication with commercial ones, as well as various
proof-of-concept and standardization efforts. They can be developed and fixed
faster than commercial ones.

Similar to domestic information security policies in many other countries, it is
common practice in Russia to specify GOST usage in various ubiquitous protocols.
These are usually just adaptations of pre-existing Russian internal standards to IETF requirements.
Examples of these GOST flavors include
\rfc{4491} for X.509,
\rfc{4490} for Cryptographic Message Syntax (CMS),
\rfc{5933} and a draft\footurl{https://tools.ietf.org/html/draft-ietf-dnsop-rfc5933-bis-00} for DNSSec,
a draft\footurl{https://tools.ietf.org/html/draft-smyslov-ike2-gost-03} for IPSec's Internet Key Exchange (IKEv2),
and drafts for TLS 1.2\footurl{https://tools.ietf.org/html/draft-smyshlyaev-tls12-gost-suites}
and TLS 1.3\footurl{https://tools.ietf.org/html/draft-smyshlyaev-tls13-gost-suites} cipher suites.

\Paragraph{GOST: digital signatures}
Informally speaking and aligning with our previous notation, the
Russian signature algorithm formula is
\begin{equation} \label{eq:gost:sign}
r = ([k]g)_x \bmod q, \quad s = k \hat{m} + r \alpha \bmod q
\end{equation}
where $\alpha$ is the signer's secret key, $k$ is
a nonce chosen randomly and uniformly from $\{1, \ldots, q-1 \}$,
$g$ is the base point of an elliptic curve, and $q$ is the order of $g$.
The GOST signature successfully verifies if
\(z_1 = s \cdot \hat{m}^{-1} \bmod q\) and \(z_2 = -r \cdot \hat{m}^{-1} \bmod q\) satisfy
\begin{equation} \label{eq:gost:verify}
([z_1]g + [z_2]P)_x = r \bmod q.
\end{equation}

In connection with these standards a number of sub-ordinary standards were
adopted (the Russian standardization system has
different levels of standards, but the difference is rather bureaucratic than
practical). In parallel the corresponding RFCs were published, including
several curves for use in GOST digital signature
algorithms. The first three curves with $\lg(p) = 256$ were described in \rfc{4357} (peculiar
that for many years it was the only normative reference to these curves---their
first appearance in Russian standards was in 2019). All these curves have only the
trivial cofactor $h=1$, i.e.\ they are cyclic groups and all curve points can be a
legal public key. After the adoption of the new digital signature standard, two
curves with $\lg(p)=512$ and $h=1$ were standardized as well as two Twisted
Edwards curves with $h=4$: one with $\lg(p) = 256$ and the other with
$\lg(p)=512$, all described in \rfc{7836}.
One important aspect is---at the standardization level---the Twisted Edwards
curves are still specified as short curves for compatibility reasons.

\Paragraph{GOST: key generation}
Russian standards say nothing about the generation of secret keys: any random
number $\alpha$ between $1$ and $q-1$ can be used as a secret key. Surprisingly
despite the Russian regulation authority paying great attention to random number
generation, there is no standard for this procedure, only some classified
requirements. The public key for a given secret one is calculated as the result
of multiplication of the curve base point by the secret key. In that sense, it
does not differ from other standard definitions of ECC key generation.

\Paragraph{GOST: key agreement (VKO)}
The VKO algorithm is defined in one of the sub-ordinary standards and described
in \rfc{7836}. It consists of 2 steps:
(i) a curve point $K$ is calculated by the formula
\[
K=[h \cdot (\mathit{UKM} \cdot x \bmod q)]Y
\]
where $x$ is the secret key of one side, $Y$ is the public key of the other
side, $\mathit{UKM}$ is an optional non-secret parameter (\textit{User Key Material})
known by both sides, $q$ is the order of the base point and $h$ is the cofactor
of the used elliptic curve;
(ii) the shared key is the hash of the affine coordinates of $K$.
In this light, VKO shares similarities to ECC CDH \cite{temp:sp:800-56ARev3}, also featuring cofactor
clearing but additionally utilizing $\mathit{UKM}$. But in contrast to NIST SP 800-108
\cite{temp:nist:sp800-108} that accounts for (the equivalent of) $\mathit{UKM}$ in the
subsequent key derivation hash function, VKO encorporates $\mathit{UKM}$ directly at the
ECC level.

\Paragraph{GOST: public key encryption}
Incorrect phrase \textit{encryption according to GOST R 34.10} is often used,
but actually the asymmetric key encryption has never been used. Instead VKO
calculates a shared key, then a symmetric encryption algorithm uses the key for
data encryption. In this light, it is \textit{hybrid encryption}.

\subsection{The GOST OpenSSL Engine} \label{sec:gostengine}

The GOST Engine project was started during OpenSSL 1.0 development. Before
OpenSSL 1.0 (released 2010) the engine mechanism allowed to provide own digests, ciphers, random
number generators, RSA, DSA, and EC. Since OpenSSL 1.0 it became possible to use
OpenSSL's engine mechanism \cite{DBLP:conf/secdev/TuveriB19}
to provide custom asymmetric algorithms.

In short, \gostengine was created as a reference implementation of the Russian GOST
cryptographic algorithms: symmetric cipher GOST 28147-89, hash algorithm GOST R
34.11-94, and asymmetric algorithms GOST R 34.10-94 (DSA-like, now deprecated
and removed) and GOST R 34.10-2001 (ECDSA-like). In 2012, the support of new
Russian hash algorithm GOST R 34.11-2012 Streebog (\rfc{6986}) and GOST R
34.10-2012 asymmetric algorithms (256 and 512 bits) was provided. After
publishing \rfc{7836} and providing non-trivial cofactor support in
OpenSSL, the support of the new parameters based on Twisted Edwards curves was
added, though the implementation itself does not use Edwards representation
and relies on OpenSSL's EC module for the curve arithmetic. It
is worth mentioning that all the parameter sets (curves) specific for GOST R
34.10-2001 are allowed for use in GOST R 34.10-2012, though the hash
algorithms are different.

Being OpenSSL-dependent software, \gostengine has been used many times as
regression testing. Not only for general engine functionality, but also for
lower level OpenSSL internals such as the EC module as discussed later in
\autoref{sec:ecckat}.

\Paragraph{Deployments}
Until OpenSSL 1.1.0 (released 2016), \gostengine was a part of OpenSSL and was distributed together.
During 1.1.0 development, the engine code was moved to a separate GitHub
repository\footurl{https://github.com/gost-engine/engine}. Currently, the engine is available as separate package in RedHat-based
Linux distributions, Debian-based distributions, and popular in the Russian ALT
Linux distribution. It is also widely used as an FOSS solution
when there is no necessity to use the officially certified
solutions. In these cases, \gostengine is often built from source instead of using the
distribution-provided packages.

\Paragraph{Asymmetric algorithms: architecture}
Asymmetric algorithm architecture in OpenSSL requires providing two opaque
callback structures per algorithm:
(i) \code{EVP\_\-PKEY\_\-ASN1\_\-METHOD} is a structure which holds a set of ASN.1
conversion, printing and information methods for a specific public key
algorithm;
(ii) \code{EVP\_\-PKEY\_\-METHOD} is a structure which holds a set of methods for a
specific public key cryptographic algorithm---those methods are usually used to
perform different jobs, such as generating a key, signing or verifying,
encrypting or decrypting, etc.
Unfortunately, because of 15-year history of the engine, the naming of the
callbacks is not extremely consistent.

\Paragraph{Asymmetric algorithms: operations}
GOST asymmetric algorithms support the following operations:
(i) key generation;
(ii) digital signature and verification;
(iii) key derivation;
(iv) symmetric 32-bytes cipher key wrap/unwrap (named encrypting/decrypting).

The best starting point is the \code{register\_\-pmeth\_\-gost} function in the
\code{gost\_\-pmeth.c} file. This function provides the setting of all the
necessary callbacks for various asymmetric algorithms. Most functions are very
similar and just call a shared wrapper around OpenSSL's EC module for the elliptic curve
arithmetic with different parameters such as hash function identifier or key length.

The following functions are especially worth studying.
(i) \code{gost\_\-ec\_\-keygen} in \code{gost\_\-ec\_\-sign.c} is the common
function for key generation, generating a random BIGNUM in the range
corresponding to the order of the selected curve's base point and calculating the
matching public key value.
(ii) \code{gost\_\-ec\_\-sign} in \code{gost\_\-ec\_\-sign.c} is the common function
for digital signature according to \rfc{7091}.
(iii) \code{gost\_\-ec\_\-verify} in \code{gost\_\-ec\_\-sign.c} is the common
function for digital signature verification according to \rfc{7091}.
(iv) \code{pkey\_\-gost\_\-ec\_\-derive} in \code{gost\_\-ec\_\-keyx.c} is the common
function for shared key derivation. This function allows two mechanisms for
derivation. The one named VKO was originally specified in \rfc{4357}, deriving
32-bytes shared key, is implemented in the \code{VKO\_\-compute\_\-key} function
in the same file. \rfc{7836} defines the other one deriving 64-bytes key using
\code{VKO\_\-compute\_\-key} as a step of key derivation. Currently, the choice
of the expected result is done by the length of a protocol-defined $\mathit{UKM}$ parameter.
(v) \code{pkey\_\-gost\_\-encrypt} in \code{gost\_\-ec\_\-keyx.c} is the common
function for symmetric key wrap using the shared key derived via
\code{pkey\_\-gost\_\-ec\_\-derive}. The key wrap for GOST 28147-89 symmetric cipher is
done according to \rfc{4357}. The key wrap for GOST R 34.12-2015 ciphers
(Kuznyechik, \rfc{7801} and
Magma\footurl{https://tools.ietf.org/html/draft-dolmatov-magma-06}) is done
according to \rfc{7836}.
(vi) \code{pkey\_\-gost\_\-decrypt} in \code{gost\_\-ec\_\-keyx.c} is the common
function for symmetric key unwrap using the shared key derived via
\code{pkey\_\-gost\_\-ec\_\-derive}. It is a reverse function for the
\code{pkey\_\-gost\_\-encrypt} function.

To summarize, regarding GOST-related ECC standards, \gostengine utilizes
OpenSSL's engine framework to its fullest: supporting key generation, key
agreement (derive in OpenSSL terminology), digital signatures and verification,
and hybrid encryption/decryption. It supports all curves from the relevant
RFCs---all the way from the test curve, to the $h=1$ short curves, to the $h=4$
short curves with Twisted Edwards equivalence. In total, eight distinct curves
with several Object Identifier (OID) aliases at the standardization level.
\section{ECC Unit Testing: ECCKAT} \label{sec:ecckat}
In this section, we present ECCKAT: a library-agnostic unit and regression
testing framework for ECC implementations. The motivation for ECCKAT began with
significant restructuring of OpenSSL's EC module introduced with major release
1.1.1 (released 2018). While the library featured simple positive testing of higher-level
cryptosystems such as ECDH and ECDSA, this provides very little confidence in
the underlying ECC implementation. To see why this is so, consider a scalar
multiplication implementation that returns a constant: this will always pass
ECDH functionality tests because the shared secret will be that constant, but is
clearly broken. Similarly on the ECDSA side, consider a verification
implementation that always returns true: this will always pass positive tests,
but is clearly broken.

With that in mind, ECCKAT uses a data-driven testing (DDT) approach heavily
relying on Known Answer Tests (KATs). The high level concept is as follows:
(i) collect existing KATs from various sources such as standards, RFCs, and
validation efforts;
(ii) augment these with negative tests and potential corner cases, and extend to
arbitrary curves using an Implementation Under Test (IUT) independent implementation;
(iii) output these tests in a standardized format, easily consumable downstream
for integration into library-specific test harnesses.
Given the wide range of curves in scope, this should be as automated as
possible. In the following sections, we expand on these aspects which make up
our implementation of ECCKAT.

\subsection{Collecting Tests}
The purpose of this first step is to build a corpus of KATs that are already
present in public documents. The goal is not only to utilize these tests but
also understand their nature, limitations, and how they can be expanded.

\Paragraph{Tests: ECC CDH}
The NIST Cryptographic Algorithm Validation Program
(CAVP)\footurl{https://csrc.nist.gov/projects/cryptographic-algorithm-validation-program}
provides test vectors for cofactor Diffie-Hellman on the following curves:
\curve{P-192}, \curve{P-224}, \curve{P-256}, \curve{P-384}, \curve{P-521},
\curve{B-163}, \curve{B-233}, \curve{B-283}, \curve{B-409}, \curve{B-571},
as well as the Koblitz curve variant of each binary curve. The test vectors
include the following fields:
\code{dIUT}, the IUT's ephemeral private key;
\code{QIUTx}, \code{QIUTy}, the IUT's ephemeral public key;
\code{QCAVSx}, \code{QCAVSy}, the peer public key;
\code{ZIUT}, the resulting shared key---in this case the $x$-coordinate of the
ECC CDH computation.
We added functionality to ECCKAT that parses these test vectors and makes them
part of the unit test corpus.

\Paragraph{Tests: ECDSA}
CAVP also provides ECDSA test vectors for the aforementioned curves, that in
fact aggregate many types of tests.
Public key validation vectors give both negative and positive tests for
(\code{Qx}, \code{Qy}) point public keys. Negative tests include coordinates out
of range (i.e.\ must satisfy $[0,p)$ for prime curves or sufficiently small
polynomial degree for binary curves) and invalid point (i.e.\ must satisfy the
curve equation), anything else being a positive test.
The negative tests are conceptually similar to the Project
Wycheproof\footurl{https://github.com/google/wycheproof} ECC-related KATs.
Key generation vectors include a private key \code{d} and the resulting
(\code{Qx}, \code{Qy}) public key point, using the default generator point.
Finally, on the ECDSA side the signing vectors include the long term private key
(\code{d}), corresponding public key (\code{Qx}, \code{Qy}), the message to be
signed (\code{Msg}), the ECDSA nonce (\code{k}), and the resulting signature
(\code{R}, \code{S}). Each test is additionally parameterized by the particular
hash function to apply to \code{Msg}.
The ECDSA verification vectors are similar, but omit the private information
\code{d} and \code{k}, also extending to both positive and negative tests
(modifying one of \code{Msg}, \code{R}, \code{S}, or the public key).
We added functionality to ECCKAT that parses these test vectors and makes them
part of the unit test corpus.

\Paragraph{Tests: Deterministic ECDSA}
The CAVP ECDSA signing tests must parameterize by the nonce to counteract the
non-determinism in stock ECDSA. In contrast, \rfc{6979} proposes a deterministic
form of ECDSA that, at a high level, computes the nonce as a function of the
private key and message to be signed. The document provides test vectors for the
exact same set of curves used in the NIST CAVP, spanning both deterministic
ECDSA signing as well as key generation.
We added functionality to ECCKAT that parses these test vectors and makes them
part of the unit test corpus. Deterministic ECDSA will likely feature in the
upcoming renewed FIPS 186-5 \cite{temp:fips:186-5}.

\subsection{Augmenting Tests}
Based on the previously collected tests and our analysis of them, the next step
is to expand these tests in several directions. First and foremost, the scope of
ECCKAT is much wider: the handful of curves above is insufficient. We extended
to general (legacy) curves over both prime and binary fields by utilizing the
SageMath computer algebra system\footurl{https://www.sagemath.org/}. This gives us an IUT-independent ground truth
during test generation. We built a large database of standardized curves with
their specific curve parameters (semi-automated with the OpenSSL \code{ecparam}
tool, listing over 80 standardized named curves), stored in JSON format that
ECCKAT parses and uses the SageMath EC module to instantiate these curves given
their parameters.

In terms of methodology, we deemed the previously collected ECDSA and
deterministic ECDSA tests sufficient. In this case, ECCKAT simply extends
coverage by allowing any legacy curve, computing the expected ECDSA output with
SageMath arithmetic. We treat key generation tests similarly, again simply
computing scalar multiplications with SageMath.

Methodology-wise, the most significant deficiency we discovered was the lack of
negative tests for ECC CDH. The reason ECC CDH differs from classical
Diffie-Hellman is to make sure the key agreement protocol fails for points of
small order in adversarial settings. Yet surprisingly none of the existing tests
actually check for this. For curves of prime order, the check is implicit
because ECC CDH and classical ECDH are equivalent. But all binary curves
(naturally including those in the original tests) have non-trivial cofactors by
definition, as well as all legacy curve equivalents of Edwards curves, Twisted
Edwards curves, and Montgomery curves require $h \geq 4$ (not in scope of the
original tests). It is a rather peculiar dichotomy since binary curves have
mostly fallen out of use, while current ECC trends for prime curves are strongly
towards these modern forms (e.g.\ both X25519 \cite{DBLP:conf/pkc/Bernstein06}
and X448 \cite{temp:x448} are standardized in \rfc{7748} and widely deployed
with e.g.\ codepoints in both TLS 1.2 \rfc{8422} and TLS 1.3 \rfc{8446}).

When applicable, i.e.\ curves with $h \neq 1$, ECCKAT generates negative tests
for ECC CDH as follows. First, with SageMath find either a generator of the full
elliptic curve group, i.e.\ an order-$hq$ point if the group is cyclic, or with
maximal order in the (in practice, rare) non-cyclic case. Scalar multiplication
by $q$ then yields a malicious generator of the largest small subgroup. This is
precisely the peer point that should produce ECC CDH protocol failure, since the
cofactor clearing (i.e.\ integer multiplication between the scalar and $h$) will
cause the resulting scalar multiplication to yield \(\mathcal{O}\): the
peer point has either order $h$ (cyclic case) or some divisor of $h$ (non-cyclic
case).

Lastly, we do note a slight deficiency in the original public key validation
negative tests. They are only \textit{partial} public key validation in that
positive tests only ensure coordinates are in range and satisfy the curve
equation. For prime-order curves, this is enough to guarantee order-$q$ points
and \textit{full} public key validation is implicit. But this is not true for
curves with $h \neq 1$. We claim this is only a minor issue because it is rare
for real-world implementations to carry out explicit \textit{full} public key
validation (i.e.\ checking that scalar multiplication by $q$ yields
\(\mathcal{O}\)) at all, since it is costly and normally handled in other more
efficient ways at the protocol level (e.g.\ with cofactor clearing).

We also added selective important corner cases for key generation. These include
positive tests for extreme private keys (i.e.\ all keys in $[1,2^b)$ and $[q - 2^b, q)$
for some reasonable bound $b > 1$) and negative tests for out of range keys
(e.g.\ negative, zero, $q$ or larger). These are important because underlying
scalar multiplication implementations often make assumptions about scalar ranges
that may or may not be ensured higher in the call stack.

We feel that such augmentation is similar (in spirit) to the work of
\citet{DBLP:conf/ctrsa/MouhaC20}, that extended NIST CAVP tests to larger
message lengths and led to \CVE{2019-8741}.

\subsection{Integrating Tests}
With the now expanded tests, the next step is applying these tests to specific
libraries. The end goal is not a one-off evaluation, but rather the ability to
apply these tests in a CI setting in an automated way and ease the integration
of these unit tests into downstream projects. To that end, we now describe three
backends ECCKAT currently supports.

\Paragraph{Test Anything Protocol (TAP)}
Our most generic solution drives TAP\footurl{http://testanything.org/} test
harnesses. With roots in Perl going back to the 80s, TAP has evolved into a
programming language-agnostic software testing framework made up of test
producers and consumers. For this backend, ECCKAT generates shell-based tests
using the Sharness\footurl{https://github.com/chriscool/sharness} portable shell
library, originally developed for Git CI. The advantage of this backend is its
portability and flexibility. The disadvantage is, while the TAP tests themselves
are library-agnostic, the test harnesses are indeed library-specific. This means
downstream projects must either parse the TAP tests themselves and convert them
to a format their internal testing framework understands (worst case), or write
simple (again, library-specific) test harness applications that conform to the
input and output expectations of the sample harnesses.

\Paragraph{OpenSSL's testing framework}
Following \CVE{2014-0160} ``Heart\-Bleed'', OpenSSL's testing framework was
rapidly overhauled and continues to evolve daily. In the scale of OpenSSL
testing (which is mostly TAP-based), the types of tests ECCKAT produces are very
low level for OpenSSL, which is much more than a cryptography library. A
significant change introduced in OpenSSL 1.1.0 (2016)---which, for the library,
marked the switch from transparent to opaque structures---expanded the
\code{evp\_\-test} harness to generically support public key operations through
OpenSSL's high level EVP (``envelope'') interface. This is precisely the correct
level to integrate ECCKAT tests.

Our OpenSSL backend for ECCKAT first encodes both private and public keys to the
PEM standard format. It does this using the \code{asn1parse} OpenSSL CLI utility
that, at a high level, directly injects ground truth ECCKAT values into an ASN1
structure, that can then be coerced to one of several portable formats---PEM in
this case, but DER is equally feasible.

In terms of test types that \code{evp\_\-test} understands, the format is a
fairly simple text file containing PEMs for key material and then test
parameters, either positive or negative; in the context of ECCKAT it supports
testing key generation, key agreement (derivation in OpenSSL terminology), and
digitally signing and verifying.

Our deterministic ECDSA KATs integrate smoothly into \code{evp\_\-test} yet
classical ECDSA does not. This issue is not OpenSSL-specific: it is normal for
libraries to handle nonce selection internally and not expose this to
application developers.
While it is possible for a library to allow overriding the ephemeral key generation,
it broadens the attack surface and introduces a potential footgun for application developers.
We feel this is strong motivation for libraries to
migrate to deterministic ECDSA, where this KAT-style testing is very natural.

The motivation for the OpenSSL backend to exist at all and not simply use a
generic TAP harness is the breadth of OpenSSL testing. A strong point of
\code{evp\_\-test} is it requires only a single application invocation, whereas
ECCKAT's generic TAP backend uses a single test harness invocation per test.
This would be prohibitively slow in the context of OpenSSL, which undergoes
rapid development and already has a significant CI load, with frequent timeouts
for GitHub PRs. In summary, test efficiency can be a practical issue, depending
on the library.

\Paragraph{GOST Engine's testing framework}
Part of the existing \gostengine test framework is Perl TAP-driven, and this is
a convenient place to integrate our ECCKAT tests. The engine already supports
key generation through the OpenSSL CLI \code{genpkey} utility. As part of our
work, as an FOSS contribution we extended \gostengine to support CLI key
agreement through the OpenSSL \code{pkeyutl} utility. At a high level, our
\gostengine backend is quite similar to the OpenSSL backend---similarly encoding
ground truth key material from ECCKAT with the \code{asn1parse} utility. The
differences are the test data being embedded directly into the Perl source as a
hash structure instead of a standalone text file to match the current test framework, and
the test logic calling the relevant OpenSSL CLI utilities to form the test
harness itself. Our ECCKAT \gostengine backend does not support GOST
digital signatures at this time. We are currently discussing porting the deterministic ECDSA
concept to GOST-style signatures. In summary, our ECCKAT \gostengine
backend provides positive and negative test coverage over all GOST curves for
both key generation and key agreement.

\subsection{ECCKAT: Results}
While we have applied and deployed ECCKAT to ECCKiila (discussed later in
\autoref{sec:ecckiila}) in a CI environment, here we
summarize our results of applying ECCKAT to other libraries. This demonstrates
the flexibility and applicability of ECCKAT.

\Paragraph{OpenSSL: ECC scalar multiplication failure}
Applying ECCKAT to \gostengine, we identified cryptosystem failures for the
\curve{id\_\-Gost\-R3410\_\-2001\_\-Crypto\-Pro\_\-C\_\-Param\-Set} curve.
Investigating the issue, OpenSSL returned failure when attempting to serialize
the output point of scalar multiplication, which was incorrectly \(\mathcal{O}\).
Internal to the OpenSSL EC module, this was due to the chosen
ladder projective formulae \cite[Eq.\ 8]{DBLP:conf/pkc/IzuT02} being undefined
for a zero $x$-coordinate divisor---a restriction noted by neither the authors nor
EFD\footurl{https://hyperelliptic.org/EFD/g1p/auto-shortw-xz.html\#ladder-ladd-2002-it-3}.
This caused the entire scalar multiplication computation to degenerate and
eventually return failure at the software level.

Broader than GOST, the $x=0$ case can happen whenever prime curve coefficient $b$ is a
quadratic residue, and we integrated this test logic into ECCKAT for all curves;
but the discovery was rather serendipitous. Most GOST curves choose the
generator point as the smallest non-negative $x$-coordinate that yields a valid
point---in this case, $x=0$. Luckily we identified this issue during the
development of OpenSSL 1.1.1, hence the issue did not affect any release version
of OpenSSL. We developed the fix for OpenSSL (\opensslpr{7000}, switching to
\cite[Eq.\ 9]{DBLP:conf/pkc/IzuT02}) as well as integrated our relevant tests
into their testing framework.

\Paragraph{OpenSSL: ECC CDH vulnerability}
Applying ECCKAT to the development branch of OpenSSL 1.1.1 identified negative
test failures in cofactor Diffie-Hellman. Investigating the issue revealed the
cause to be mathematically incorrect side channel mitigations at the scalar
multiplication level. As a timing attack countermeasure (ported from
\CVE{2011-1945} by \cite{DBLP:conf/esorics/BrumleyT11}), the ladder code first
padded the scalar by adding either $q$ or $2q$ to fix the bit length and
starting iteration of the ladder loop. But
in key agreement scenarios, there is no guarantee the peer point is an order-$q$
point---only a point with order dividing $hq$ if it satisfies the curve
equation, i.e.\ is an element of the elliptic curve group. This caused negative
tests to fail for all curves with a non-trivial cofactor---for named curves in
OpenSSL, this included all binary curves and the 112-bit \curve{secp112r2}
Certicom curve with $h=4$.

Luckily the issue did not affect any release version of OpenSSL. We developed
the fix for OpenSSL (\opensslpr{6535}) as well as integrated our relevant tests
into their testing framework.
In support of Open Science, we released a demo for this vulnerability as a
research artifact \cite{zenodo:2020:acsac}.

\Paragraph{GOST: VKO vulnerability}
Applying ECCKAT to \gostengine identified negative test failures in VKO key
agreement for the two curves with non-trivial cofactors, similar (in spirit) to
the cofactor Diffie-Hellman failures above. Investigating the issue revealed
\gostengine multiplied by the cofactor \textit{before} modular reduction.
Consulting the Russian standard and \rfc{7836}, surprisingly this is in fact a
valid interpretation of VKO at the standardization level.

Prior to the standard change and RFC errata resulting from our work, both the
Russian standard and RFC specified VKO computation\footnote{Here the scalar
multiplication notation is per the RFC to make the errata clear.} as
\[
(m/q \cdot \mathit{UKM} \cdot x \bmod q) \cdot (y \cdot P)
\]
where, recalling from \autoref{sec:gostmath}, $m=hq$ is the curve cardinality,
$\mathit{UKM}$ is user key material, $x$ is the private key, $q$ is the order of the
generator, and $y \cdot P$ is the peer public key (point). With this
formulation, the cofactor clearing is ineffective: it is absorbed modulo $q$.
For the two curves satisfying $h=4$, in case of a malicious $y \cdot P$ such as
an order-$h$ point, the computation results in one of the four points in the
order-$h$ subgroup, i.e.\ a small subgroup confinement attack. This can reveal
the private key value modulo $h$ and, depending on the protocol, force session
key reuse.

Subsequent to our work, the Russian standard and \rfc{7836} now specify the compatible (in the
non-adversarial sense)
\[
(m/q \cdot (\mathit{UKM} \cdot x \bmod q)) \cdot (y \cdot P)
\]
where it is explicit the cofactor clearing is \textit{after} the modular
reduction. As part of our work, we implemented the \gostengine fix
(\href{https://github.com/gost-engine/engine/pull/265}{PR \#265}) and integrated
all the relevant ECCKAT positive and negative tests into the \gostengine testing
framework. Luckily, packaged versions of \gostengine for popular distributions
such as Debian, Ubuntu, and RedHat use older versions of the engine that only
feature the $h=1$ curves, not affected by this vulnerability.
\section{Generating ECC Layers: ECCKiila} \label{sec:ecckiila}
\begin{figure}
\centering
\includegraphics[width=1.0\linewidth]{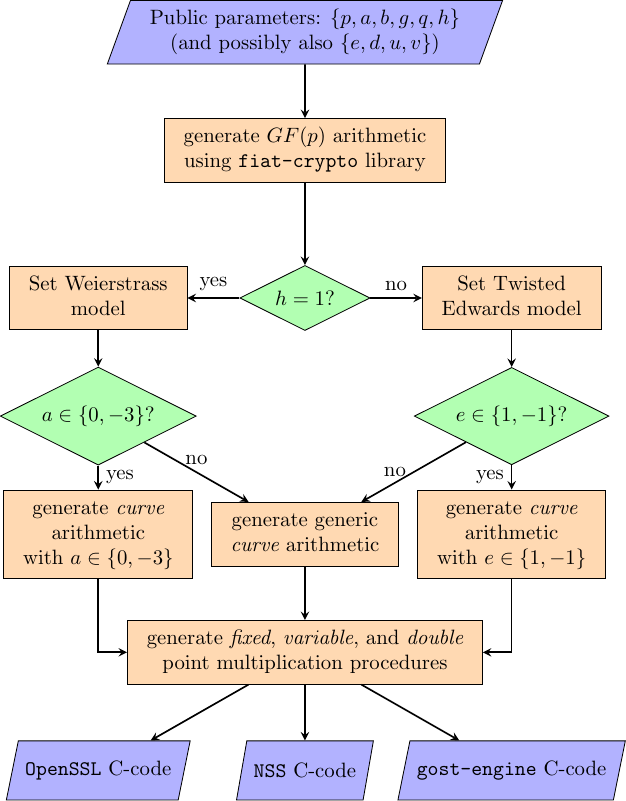}
\caption{General concept of ECCKiila. The public parameters determine a
Weierstrass curve \(E_w \colon y^2 = x^3 + ax + b\) such that \(\#E_w = h\cdot
q\) and \(g\) is an order-\(q\) {\it affine} point. The optional values
determine the Twisted Edwards curve \(E_t \colon eu^2 + v^2 = 1 + du^2v^2\) and
the image point \((u,v)\) of \(g\) on \(E_t\).}
\label{fig:tikz:dflow}
\end{figure}

This section focuses on the ECC layer generation and required library-specific
rigging. \autoref{fig:tikz:dflow} summarizes our proposed full stack
implementation named \textit{ECCKiila}. The name comes from the Finnish word
{\it kiila} that means {\it wedge}, and it allows to dynamically create the
C-code (supporting both 64-bit and 32-bit architectures, no alignment or
endianness assumptions) regarding to the ECC layer as well as the rigging for
seamless integration into {\tt OpenSSL}, {\tt NSS}, and \gostengine, all driven
by Python Mako templating. \autoref{tb:curves} shows all the curves tested with
ECCKiila.
To establish scope, while timing attacks are included in the ECCKiila threat model,
physical (e.g.\ power, electromagnetic emanations) or invasive (e.g.\ fault)
attack techniques are not.

\Paragraph{Field arithmetic}
In our proposal, we obtain the majority of the $GF(p)$ arithmetic by using the
{\tt fiat-crypto} project\footurl{https://github.com/mit-plv/fiat-crypto}, that
provides generation of field-specific formally verified constant time code
\cite{DBLP:conf/sp/ErbsenPGSC19}. \texttt{Fiat-crypto} has several strategies to generate 
the arithmetic but we have chosen the best per curve base on the form of $p$. 
In other words, the remainder of the section is centered on the EC layer 
that builds on top of the $GF(p)$ layer. It is important to note this is 
the formal verification boundary for ECCKiila---all other code on top of Fiat, 
while computer generated through templating and automatic formula generation 
for ECC arithmetic, has no formal verification guarantees.
From now on, we assume all operations are performed in \(GF(p)\).

\Paragraph{Short Weierstrass curves}
All the legacy curves we consider in this work are prime order
curves \(E_w\), i.e.\ \(\#E_w = q \approx p\) is a prime number and $h=1$.

\Paragraph{Twisted Edwards curves}
Recall from \autoref{sec:gostengine} most of the legacy curves from \gostengine
work on curves \(E_w\) of prime cardinality \(q \approx p \) but two of them are
centered on curves of cardinality \(4q\) being \(q \approx p\) a prime number.
For those two special curves, \gostengine curves are represented in short
Weiestrass form at the specification level (i.e.\ ``on-the-wire'' or when
serialized) but internally we use the Twisted Edwards curve representation.
Additionally, we implemented the mappings that connect \(E_w\) and \(E_t\) by
writing \eqref{sec:ecckiila:eq1} and \eqref{sec:ecckiila:eq2} in their
{\it projective} form to delay the costly inversion in $GF(p)$.
We used the same strategy for \curve{MDCurve20160}, the
``Million Dollar Curve'' \cite{temp:mdc} as a research-oriented example.
\Paragraph{Point arithmetic}
The way of adding points depends on the curve model being used, but we describe
the three main point operations as follows: (i) the {\it mixed point addition}
that takes as inputs a {\it projective} point and an {\it affine} point, and it
returns a {\it projective} point; (ii) the {\it projective point addition} and
(iii) the {\it projective point doubling}, which their inputs and outputs are
{\it projective} points.

We use the {\it exception-free} formulas proposed by
\citet[Sec.\ 3]{DBLP:conf/eurocrypt/RenesCB16} and
\citet[Sec.\ 3.1, Sec.\ 3.3]{DBLP:conf/asiacrypt/HisilWCD08}
for Weierstrass and Twisted Edwards models, respectively. In particular, all
ECC arithmetic is machine generated, tied to the \code{op3}
files\footnote{see e.g.\
\url{https://www.hyperelliptic.org/EFD/g1p/auto-code/twisted/extended/doubling/dbl-2008-hwcd.op3}}
included in our software implementation. Our tooling is configurable in that
sense, but also with high correctness confidence.

Now, recall in Weierstrass models \(\mathcal{O}\) has no {\it affine}
representation and thus the {\it mixed point addition} needs to catch whether
the {\it affine} point describes \(\mathcal{O}\). We solve this by asking if its
affine \(Y\)-coordinate is zero and performing a conditional copy at the end of
the {\it mixed point addition}, all in constant time. That is, we set \((0,0)\)
as the {\it affine} representation of \(\mathcal{O}\), which does not satisfy
the {\it affine} curve equation of \(E_w\) (no order-2 point exists on curves
with prime cardinality, hence $y=0$ is a contradiction). However, this is not
the case for Twisted Edwards models, which allow fully {\it exception-free}
formulas for point addition procedures.

\Paragraph{Point multiplication}
The heart of our ECC layer is point multi\-plication---\textit{fixed} point
\(g\), \textit{variable} point \(P\), and also by the \textit{double} point
multiplication \([k]g + [\ell]P\).
We implemented the \textit{variable} point multiplication by representing
scalars with the regular-NAF method \cite[Sec.\ 3.2]{DBLP:conf/africacrypt/JoyeT09}
and $\lg(q)/w$ digits with $\lg(q)$ doublings. The advantage of this method is
we need only half the precomputed values compared to e.g.\ the base $2^w$
method. We support a variable window length $w$, and by default $w=5$.
We implemented the \textit{fixed} point multiplication using the \textit{comb}
method (see \cite[9.3.3]{DBLP:reference/crc/2005ehcc}) with interleaving and,
similar to the variable point case, using the regular-NAF scalar representation.
Our approach seeks full generality and word size independence on each
architecture (32 or 64 bit), hence we automatically calculate the number of
\textit{comb}-teeth $\beta$ and the distance between consecutive teeth
$\lceil \lg(q)/\beta \rceil$, where the latter should be a multiple of $w$,
considering the size of the L1 cache in this process. Therefore, the static LUTs
span $\beta$ tables requiring $\lceil \lg(q)/\beta\rceil$ doublings.
Both methods are constant time, performing exactly $\lg(q)/w$ point additions,
and using linear passes to ensure key-independent memory access not only to LUTs
but also in conditional point negations due to the signed scalar representation
and conditional trailing subtraction to handle even scalars; the regular-NAF
encoding itself is also constant time.
We implemented \textit{double} point multiplication using textbook $w$NAF
\cite[9.1.4]{DBLP:reference/crc/2005ehcc} combined with Shamir's trick
\cite[9.1.5]{DBLP:reference/crc/2005ehcc}. This shares the doublings, i.e.\
maximum $\lg(q)$ in number, but on average reduces the number of additions per
scalar. This is because it is variable-time---only required in digital
signature verification where all inputs are public.

\Paragraph{Rigging}
At this point, the resulting C code yields functional arithmetic for the ECC
layer. But we observe a gap between such code and real world projects. On one
hand, researchers intimately familiar with ECC details lack the skill,
motivation, and/or domain-specific knowledge to integrate the ECC stack into
downstream large software projects. On the other hand, developers for those
downstream projects lack the intimate knowledge of the upstream ECC layer to
integrate properly---historical issues include assumptions on architecture,
alignment, endianness, supported ranges to name a few. For example, one obscure
issue encountered during OpenSSL ECCKiila integration was lack of OpenSSL unit
testing for custom ECC group base points, which OpenSSL supports but ECCKiila
cannot fully accelerate since the generated LUTs are static, but regardless must
detect and handle the base point mismatch. Our integrations passed all OpenSSL
unit tests, which is clearly not correct in this corner case.

To solve this issue, the last layer of ECCKiila is \textit{rigging} that is
essentially plumbing for downstream projects. Rigging is library-specific by
nature, and ECCKiila currently supports three backends: OpenSSL, NSS, and
\gostengine. For OpenSSL, ECCKiila generates an \code{EC\_\-GROUP} structure,
which is the OpenSSL internal representation of a curve with function pointers for
various operations. We designed simple wrappers for three relevant function
pointers, which are shallow and eventually (after sanity checking arguments and
transforming the inputs to the expected format) call the corresponding scalar
multiplication implementation in \autoref{fig:tikz:dflow}. NSS is similar with
an \code{ECGroup} structure. The \gostengine rigging (mostly) decouples it from
OpenSSL's EC module, since it only needs to support GOST curves with explicit
parameters.

\begin{table}
\resizebox{0.47\textwidth}{!}{%
\begin{tabular}{|l|l||r|r|r|}
\hline
\multirow{2}{*}{Curve} & \multirow{2}{*}{Library} & External model & Internal model & \multirow{2}{*}{\(\lg (p)\)}\\
& & (Standard) & (ECCKiila) & \\
\hline
\curve{secp192r1} / \curve{P-192} & \multirow{12}{*}{OpenSSL} & \multicolumn{2}{|c|}{\multirow{2}{*}{Weierstrass with $a = -3$}} & 192 \\ \cline{5-5}
\curve{secp256r1} / \curve{P-256} & & \multicolumn{2}{r|}{} & 256 \\ \cline{5-5}\cline{3-4}
\curve{secp256k1} & & \multicolumn{2}{c|}{Weierstrass with \(a = 0\)} & 256 \\ \cline{5-5}\cline{3-4}
\curve{secp384r1} & & \multicolumn{2}{|c|}{\multirow{7}{*}{Weierstrass with $a = -3$}} & 384 \\ \cline{5-5}
\curve{secp521r1} / \curve{P-521} & & \multicolumn{2}{r|}{} & 521 \\ \cline{5-5}
\curve{brainpool192t1} & & \multicolumn{2}{r|}{} & 192 \\ \cline{5-5}
\curve{brainpool256t1} & & \multicolumn{2}{r|}{} & 256 \\ \cline{5-5}
\curve{brainpool320t1} & & \multicolumn{2}{r|}{} & 320 \\ \cline{5-5}
\curve{brainpool384t1} & & \multicolumn{2}{r|}{} & 384 \\ \cline{5-5}
\curve{brainpool512t1} & & \multicolumn{2}{r|}{} & 512 \\ \cline{5-5}
\curve{SM2} (Chinese standard) & & \multicolumn{2}{r|}{} & 256 \\ \cline{5-5} \cline{1-1}\cline{3-4}\cline{5-5}
\curve{X25519} / \curve{ED25519} / \curve{Wei25519} & & \multirow{2}{*}{\begin{tabular}{r}Weierstrass with\\ $a\neq0$ and $a\neq -3$ \end{tabular}} & Edwards with $e = -1$ & 255 \\ \cline{4-5}
\curve{X448} / \curve{ED448} / \curve{Wei448} & & & Edwards with $e = 1$ & 448 \\ \cline{3-4}\cline{5-5}
\cline{1-2}
\curve{P-384} & \multirow{2}{*}{NSS} & \multicolumn{2}{r|}{} & 384 \\ \cline{5-5}
\curve{P-521} & & \multicolumn{2}{r|}{} & 521\\ \cline{5-5}
\cline{1-2}
\curve{id\_GostR3410\_2001\_CryptoPro\_A\_ParamSet} & \multirow{8}{*}{GOST} & \multicolumn{2}{r|}{} & \multirow{3}{*}{256}\\
\curve{id\_GostR3410\_2001\_CryptoPro\_B\_ParamSet} & & \multicolumn{2}{r|}{} & \\
\curve{id\_GostR3410\_2001\_CryptoPro\_C\_ParamSet} & & \multicolumn{2}{r|}{} & \\ \cline{5-5}
\curve{id\_tc26\_gost\_3410\_2012\_512\_paramSetA} & & \multicolumn{2}{r|}{} &\multirow{3}{*}{512} \\
\curve{id\_tc26\_gost\_3410\_2012\_512\_paramSetB} & & \multicolumn{2}{r|}{} & \\ \cline{5-5}
\cline{3-4}
\curve{id\_tc26\_gost\_3410\_2012\_256\_paramSetA} & & \multirow{3}{*}{\begin{tabular}{r}Weierstrass with\\ $a\neq0$ and $a\neq -3$ \end{tabular}} & \multirow{3}{*}{\begin{tabular}{r}Twisted Edwards\\ with $e = 1$ \end{tabular}} & 256\\ \cline{5-5}
\curve{id\_tc26\_gost\_3410\_2012\_512\_paramSetC} & & & & 512\\ \cline{5-5}
\cline{1-2}
\curve{MDCurve20160} (Million Dollar Curve) & --- & & & 256\\ \cline{5-5}
\hline
\end{tabular}
}
\caption{List of all the curves tested with ECCKiila}
\label{tb:curves}
\end{table}

\Paragraph{Example: P-384 in OpenSSL and NSS}
What follows is a walkthrough of our integration of \curve{secp384r1} into
OpenSSL and NSS. ECCKiila has a large database (JSON) of standard curves, then
generates both 64-bit and 32-bit $GF(p)$ arithmetic using Fiat. It then takes
the $h=1$, $a=-3$ path in \autoref{fig:tikz:dflow} and generates the three
relevant scalar multiplication functions that utilize exception-free formulas
from \cite{DBLP:conf/eurocrypt/RenesCB16} optimized for $a=-3$. Finally,
ECCKiila emits the OpenSSL rigging for OpenSSL integration, and NSS rigging for
NSS integration. Adding the code to OpenSSL is the only current manual step: one
line to add the new code to the build system, one line to add the prototype of
the new \code{EC\_\-METHOD} structure in a header, and one line to point OpenSSL
at this structure for the \curve{secp384r1} definition. The NSS integration is
very similar.

\Paragraph{Example: GOST twisted 256-bit curve}
What follows is a walkthrough of our integration of
\curve{id\_\-tc26\_\-gost\_\-3410\_\-2012\_\-256\_\-paramSetA} into \gostengine.
ECCKiila takes the $h=4$, $e=1$ path in \autoref{fig:tikz:dflow} and generates
the three relevant scalar multiplication functions that utilize exception-free
formulas from \cite{DBLP:conf/asiacrypt/HisilWCD08} optimized for $e=1$, and
noting the $E_w$ to $E_t$ mappings (and back) are transparent to the caller
(\gostengine rigging, in this case). Finally, ECCKiila emits the \gostengine
rigging, and enabling this code in \gostengine is currently the only manual
step: one line to add the code to the \gostengine build system, and three lines
in a C switch statement to enable each of the relevant scalar multiplication
routines.

\Paragraph{Example: million dollar curve in OpenSSL}
Not to limit ECCKiila to only formally standardized curves, here we showcase the
research value of ECCKiila by applying it to \curve{MDCurve20160}
\cite{temp:mdc}, which has undergone no formal standardization process. As such,
we took the GOST approach that perhaps the $E_w$ curve might be standardized,
and the $E_t$ curve utilized internally. We applaud this approach in GOST
because, in practice, it eases downstream integration and lowers the effort bar
during standardization---on the downside, it does reduce flexibility since it
implies compliance with certain existing (legacy) standards.

The process for ECCKiila is a logical mix of the previous two examples: taking a
path similar to the GOST example, but the generated rigging is OpenSSL. In this
case, OpenSSL knows nothing about \curve{MDCurve20160} so we obtained an
unofficial OID for \curve{MDCurve20160} and the rigging additionally emits the
explicit curve parameters so OpenSSL knows how to construct its internal ECC
group. The only manual steps are similar to the previous examples, yet
additionally inserting these parameters.

Once OpenSSL knows about \curve{MDCurve20160} from the automated rigging, it can
drive operations with \curve{MDCurve20160} like any other (legacy) curve: ECC
key generation, ECDSA signing and verifying, and ECC CDH key agreement. This
highlights the research value of ECCKiila, and gives a clear and simple path for
researchers seeking dissemination and exploitation: obtain an official OID for
standardization, provide curve parameters to ECCKiila, and submit a PR to
downstream projects. In the case that more modern signature and key agreement
schemes are desired, additional steps are needed at both the standardization,
implementation, and integration levels.

\subsection{ECCKiila: Results} \label{sec:results}
We now present the results of applying ECCKiila to the curves listed in
\autoref{tb:curves}. First, it is important to note our measurements are not on
the ECCKiila code directly, rather on the application-level view of how
developers and users of the corresponding libraries will transparently see the
resulting performance difference. That is, we are measuring the \textit{full
integration}, not the ECCKiila code in isolation. So it includes e.g.\ all
overheads from the rigging, any checks and serialization/deserialization the
libraries perform, any memory allocation/deallocation and structure
initialization, as well as any other required arithmetic not part of ECCKiila
(e.g. $GF(q)$ arithmetic for ECDSA and GOST signatures).

To compare the performance of our approach, we measure the
timing of unmodified OpenSSL 3.0 alpha, \gostengine, and NSS 3.53 (called
\textit{baseline}), and the same versions then modified with the ECCKiila
output (\textit{integration}). For each one of them, we measured the timings
from the operations described in \autoref{sec:background} such as key
generation, key agreement (derivation), signing, and verification. For the sake
of simplicity, we refer to them as keygen, derive, sign, and verify,
respectively.

The hardware and software setup used to get the timings reported in this section
are the following: Intel Xeon Silver 4116 2.10GHz, Ubuntu 16.04 LTS ``Xenial'',
GNU11 C, and \code{clang-10}. We used 64-bit builds, although the ECCKiila generated
code selects the correct implementation using the compiler's preprocessor at
build time.
For the clock cycle measurements, we used the
\code{newspeed}\footurl{https://github.com/romen/newspeed} utility, unifying the
OpenSSL and \gostengine measurements since it works through OpenSSL's EVP
interface and optionally supports engines. For the three NSS results, we modified
their \code{ecperf} benchmarking
utility\footurl{https://github.com/nss-dev/nss/tree/master/cmd/ecperf} to report
median clock cycles instead of wall clock time.

\autoref{tab:bench} reports timings for both approaches, showing the result of
our proposal has good performance regarding the original versions. There are
several nuances to clarify in the data. In particular, in the signature of our
proposal where \curve{id\_\-GostR3410\_\-2001\_\-CryptoPro\_\-A\_\-ParamSet} is
quite faster than \curve{secp256k1}: the reason is GOST signatures do not invert
modulo $q$ while ECDSA signatures do, and this is a costly operation.
Also, we can see that \curve{secp256r1} has excellent performance due to
manual AVX assembler optimizations, while ECCKiila is portable C.
Despite this, \curve{id\_\-tc26\_\-gost\_\-3410\_\-2012\_\-256\_\-paramSetA}
gives us similar performance, yet completely automated with limited formal
verification guarantees and architecture independence.
Last, there are some curve with extra slowdown in some operations such as
\curve{secp256r1}, \curve{brainpoolP512t1}, and \curve{secp521r1}. The reason
for this varies. In the \curve{secp256r1} and \curve{secp521r1} cases, this is
due to competition with curve-specific optimizations in libraries. For
Brainpool curves, this is a combination of limited $GF(p)$ optimizations
available both at the {\tt fiat-crypto} and ECCKiila layers.

\begin{table*}
\centering
\resizebox{1.0\textwidth}{!}{
\begin{tabular}{|l|c|r|r||r|r||r|r||r|r|}
\hline
\multirow{2}{*}{\bf Curve/Parameter}& \multirow{2}{*}{\bf \rotatebox[origin=c]{270}{bit}} & \multicolumn{2}{ |c|| }{{\bf KeyGen}} & \multicolumn{2}{ |c|| }{{\bf Derive}} & \multicolumn{2}{ |c|| }{{\bf Sign}} & \multicolumn{2}{ |c| }{{\bf Verify}} \\ \cline{3-10}
 & & Baseline & Integration & Baseline & Integration & Baseline & Integration & Baseline & Integration \\ \hline
\curve{secp192r1} & \multirow{2}{*}{\rotatebox[origin=c]{270}{{\small 192}}} & 587  & 77   ($\blacktriangle$  7.6x) & 549  & 181  ($\blacktriangle$  3.0x) & 574  & 75   ($\blacktriangle$  7.6x) & 543  & 212  ($\blacktriangle$  2.6x) \\
\curve{brainpoolP192t1} & & 574  & 92   ($\blacktriangle$  6.2x) & 533  & 255  ($\blacktriangle$  2.1x) & 560  & 90   ($\blacktriangle$  6.2x) & 543  & 291  ($\blacktriangle$  1.9x) \\ \hline
\curve{X25519} / \curve{ED25519} / \curve{Wei25519} & \multirow{1}{*}{\rotatebox[origin=c]{270}{{\small 255}}} & 105  & 91   ($\blacktriangle$  1.2x) & 104  & 173  ($\triangledown$  1.7x) & 106  & 112  ($\triangledown$  1.1x) & 284  & 211  ($\blacktriangle$  1.3x) \\ \hline
\curve{secp256r1} & \multirow{10}{*}{\rotatebox[origin=c]{270}{{\small 256}}} & 90   & 141  ($\triangledown$  1.6x) & 139  & 465  ($\triangledown$  3.3x) & 63   & 156  ($\triangledown$  2.5x) & 183  & 524  ($\triangledown$  2.9x) \\
\curve{P-256} (NSS) & & 310  & 116  ($\blacktriangle$  2.7x) & 1628 & 916  ($\blacktriangle$  1.8x) & 351  & 157  ($\blacktriangle$  2.2x) & 1077 & 512  ($\blacktriangle$  2.1x) \\
\curve{secp256k1} & & 1027 & 151  ($\blacktriangle$  6.8x) & 989  & 400  ($\blacktriangle$  2.5x) & 1037 & 165  ($\blacktriangle$  6.3x) & 932  & 471  ($\blacktriangle$  2.0x) \\
\curve{brainpoolP256t1} & & 939  & 175  ($\blacktriangle$  5.3x) & 897  & 597  ($\blacktriangle$  1.5x) & 953  & 187  ($\blacktriangle$  5.1x) & 877  & 665  ($\blacktriangle$  1.3x) \\
\curve{id\_GostR3410\_2001\_CryptoPro\_A\_ParamSet} & & 1026 & 123  ($\blacktriangle$  8.3x) & 1022 & 385  ($\blacktriangle$  2.7x) & 993  & 91   ($\blacktriangle$ 10.8x) & 867  & 404  ($\blacktriangle$  2.1x) \\
\curve{id\_GostR3410\_2001\_CryptoPro\_B\_ParamSet} & & 982  & 129  ($\blacktriangle$  7.6x) & 1007 & 449  ($\blacktriangle$  2.2x) & 955  & 101  ($\blacktriangle$  9.4x) & 845  & 461  ($\blacktriangle$  1.8x) \\
\curve{id\_GostR3410\_2001\_CryptoPro\_C\_ParamSet} & & 977  & 180  ($\blacktriangle$  5.4x) & 986  & 639  ($\blacktriangle$  1.5x) & 945  & 145  ($\blacktriangle$  6.5x) & 848  & 662  ($\blacktriangle$  1.3x) \\
\curve{id\_tc26\_gost\_3410\_2012\_256\_paramSetA} & & 960  & 101  ($\blacktriangle$  9.5x) & 929  & 204  ($\blacktriangle$  4.5x) & 926  & 69   ($\blacktriangle$ 13.3x) & 893  & 240  ($\blacktriangle$  3.7x) \\
\curve{MDCurve201601} & & --- & 155  \phantom{!!!!!!!!!!!} & --- & 360  \phantom{!!!!!!!!!!!} & --- & 171  \phantom{!!!!!!!!!!!} & --- & 420  \phantom{!!!!!!!!!!!} \\
\curve{SM2} & & 1039 & 252  ($\blacktriangle$  4.1x) & 889  & 549  ($\blacktriangle$  1.6x) & 935  & 174  ($\blacktriangle$  5.4x) & 874  & 612  ($\blacktriangle$  1.4x) \\ \hline
\curve{brainpoolP320t1} & \multirow{1}{*}{\rotatebox[origin=c]{270}{{\small 320}}} & 1470 & 314  ($\blacktriangle$  4.7x) & 1430 & 1161 ($\blacktriangle$  1.2x) & 1502 & 351  ($\blacktriangle$  4.3x) & 1277 & 1271 ($\blacktriangle$  1.0x) \\ \hline
\curve{secp384r1} & \multirow{3}{*}{\rotatebox[origin=c]{270}{{\small 384}}} & 2156 & 417  ($\blacktriangle$  5.2x) & 2117 & 1598 ($\blacktriangle$  1.3x) & 2221 & 488  ($\blacktriangle$  4.5x) & 1818 & 1823 ($\triangledown$  1.0x) \\
\curve{P-384} (NSS) & & 2257 & 391  ($\blacktriangle$  5.8x) & 4266 & 3225 ($\blacktriangle$  1.3x) & 2310 & 454  ($\blacktriangle$  5.1x) & 4013 & 1755 ($\blacktriangle$  2.3x) \\
\curve{brainpoolP384t1} & & 2157 & 527  ($\blacktriangle$  4.1x) & 2098 & 1978 ($\blacktriangle$  1.1x) & 2206 & 599  ($\blacktriangle$  3.7x) & 1828 & 2181 ($\triangledown$  1.2x) \\ \hline
\curve{X448} / \curve{ED448} / \curve{Wei448} & \multirow{1}{*}{\rotatebox[origin=c]{270}{{\small 448}}} & 309  & 306  ($\blacktriangle$  1.0x) & 1046 & 760  ($\blacktriangle$  1.4x) & 322  & 406  ($\triangledown$  1.3x) & 1195 & 903  ($\blacktriangle$  1.3x) \\ \hline
\curve{brainpoolP512t1} & \multirow{4}{*}{\rotatebox[origin=c]{270}{{\small 512}}} & 3632 & 1325 ($\blacktriangle$  2.7x) & 3590 & 4767 ($\triangledown$  1.3x) & 3774 & 1451 ($\blacktriangle$  2.6x) & 2959 & 5099 ($\triangledown$  1.7x) \\
\curve{id\_tc26\_gost\_3410\_2012\_512\_paramSetA} & & 3716 & 664  ($\blacktriangle$  5.6x) & 3634 & 2197 ($\blacktriangle$  1.7x) & 3671 & 625  ($\blacktriangle$  5.9x) & 2878 & 2405 ($\blacktriangle$  1.2x) \\
\curve{id\_tc26\_gost\_3410\_2012\_512\_paramSetB} & & 3754 & 652  ($\blacktriangle$  5.8x) & 3720 & 2359 ($\blacktriangle$  1.6x) & 3715 & 619  ($\blacktriangle$  6.0x) & 2798 & 2550 ($\blacktriangle$  1.1x) \\
\curve{id\_tc26\_gost\_3410\_2012\_512\_paramSetC} & & 3663 & 515  ($\blacktriangle$  7.1x) & 3653 & 1262 ($\blacktriangle$  2.9x) & 3645 & 478  ($\blacktriangle$  7.6x) & 3070 & 1387 ($\blacktriangle$  2.2x) \\ \hline
\curve{secp521r1} & \multirow{2}{*}{\rotatebox[origin=c]{270}{{\small 521}}} & 574  & 513  ($\blacktriangle$  1.1x) & 941  & 1731 ($\triangledown$  1.8x) & 753  & 742  ($\blacktriangle$  1.0x) & 1492 & 2096 ($\triangledown$  1.4x) \\
\curve{P-521} (NSS) & & 3239 & 480  ($\blacktriangle$  6.7x) & 6063 & 3444 ($\blacktriangle$  1.8x) & 3339 & 578  ($\blacktriangle$  5.8x) & 5217 & 1840 ($\blacktriangle$  2.8x) \\ \hline
 \hline
\end{tabular}
}
\caption{Comparison of timings between the baseline and the integration from
OpenSSL, \gostengine, and NSS. All timings are reported in clock cycles (thousands).}
\label{tab:bench}
\end{table*}
\section{Conclusion} \label{sec:conclusion}
In this work, we presented two methodologies.
ECCKAT allows carrying out a set of tests over an arbitrary ECC implementation,
including (but not limited to) all standard curves from OpenSSL, NSS, and
\gostengine where it gave us excellent results because we uncovered several
novel defects in OpenSSL such as a scalar multiplication failure and an ECC CDH
vulnerability. Meanwhile, for GOST we detected a VKO vulnerability that can
reveal sensitive information from the private key.
Our second proposal ECCKiila is partially motivated by these vulnerabilities.
With the use of ECCKiila, we can generate code dynamically for any curve,
including all standard curves from OpenSSL, NSS, and GOST. This code is highly
competitive in comparison with the original versions from OpenSSL, NSS, and
\gostengine since we have a speedup factor up to 9.5x for key generation, 
4.5x for key agreement, 13.3x for signing, and 3.7x for verifying 
as \autoref{tab:bench} shows. Furthermore, ECCKiila is flexible and robust since 
we can easily add new curves without increasing the development complexity---upstream
or downstream. Hence, we believe our methodologies are of interest for future work, 
both in research and application.

Quoting \citet{DBLP:conf/usenix/Davis01}: \textit{programmers need ``turnkey''
cryptography, not only cryptographic toolkits} and that is precisely what
ECCKiila provides. The ease of integrating these stacks in downstream projects,
coupled with formal verification guarantees on the Galois field arithmetic and
simplicity of upper layers, and automated code generation, provides drop-in,
zero-maintenance solutions for real-world, security-critical libraries.
We release ECCKiila\footurl{https://gitlab.com/nisec/ecckiila} as FOSS,
furthermore in support of Open Science.

\Paragraph{Impact}
ACSAC 2020's Hard Topic Theme is ``Deployable and Impactful Security'' and our
work meets that challenge.
All of the defects uncovered by ECCKAT in \autoref{sec:ecckat} are fixed and
merged into the related projects.
ECCKiila stacks are already merged in two downstream projects. ECCKiila now
provides all ECC in \gostengine, and NSS merged our ECCKiila stacks for
P-384\footurl{https://hg.mozilla.org/projects/nss/rev/d19a3cd451bbf9602672fdbba8d6a817a55bfc69} and
P-521\footurl{https://hg.mozilla.org/projects/nss/rev/ca068f5b5c176c503ddce969e78dd326cc5fd29a}
as a result of \CVE{2020-6829} by \citet{temp:dejavu}.

\Paragraph{Future work}
ECCKiila itself has no formal verification guarantees, and only inherits those
provided by Fiat for the vast majority of the $GF(p)$ layer. One potential
research direction would be investigating the possibility of extending proofs to
cover more of ECCKiila, particularly at the ECC layer.

\Paragraph{Acknowledgments}
This project has received funding from the European Research Council (ERC) under
the European Union's Horizon 2020 research and innovation programme (grant
agreement No 804476).
 
\bibliographystyle{ACM-Reference-Format}

\end{document}